\documentclass[fleqn,twoside]{article}
\usepackage{espcrc2}
\usepackage{graphicx}

\usepackage{graphicx}

\newcommand{\AmS}{{\protect\the\textfont2
  A\kern-.1667em\lower.5ex\hbox{M}\kern-.125emS}}

\title{Chiral symmetry breaking in (2+1) dimensional QED}

\author{Pieter Maris\address{Dept. of Physics, 
	North Carolina State University,
        Box 8202, Raleigh,  NC 27695}%
        \thanks{Financially supported by DOE under grants
		No. DE-FG02-96ER40947 and DE-FG02-97ER41048.}
        and
	Dean Lee\addressmark
	\thanks{This work benefitted from the resources 
		of the North Carolina Supercomputer Center.}
	}
       
\begin{document}

\begin{abstract}
We study dynamical mass generation in QED in (2+1) dimensions using
Hamiltonian lattice methods.  We use staggered fermions, and perform
simulations with explicit dynamical fermions in the chiral limit.  We
demonstrate that a recently developed method to reduce the fermion
sign problem can successfully be applied to this problem.  Our results
are in agreement with both the strong coupling expansion and with
Euclidean lattice simulations.
\vspace{1pc}
\end{abstract}

\maketitle

\section{INTRODUCTION}

QED in (2+1) dimensions exhibits both dynamical mass
generation~\cite{pisa,dcsb,Burkitt:1987nx,Hamer:1997bf,Hands:2002dv}
and confinement~\cite{conf}.  This makes it an interesting theory to
study these nonperturbative phenomena.  Furthermore, these phenomena
in QED$_3$ are relevant for applications in condensed matter physics,
in particular in connection with high-$T_c$
superconductivity~\cite{superTc}.  Here, we focus on the question of
chiral symmetry breaking in QED$_3$.

We use staggered fermions, and perform simulations with explicit
dynamical fermions in the chiral limit.  Simulations with explicit
fermions could improve our understanding of the microscopic behavior
of (chiral) fermions in lattice simulations.  We apply a recently
developed method~\cite{dean02}, called the ``zone method'', to
reduce the fermion sign problem in our simulation.  Preliminary
results for the chiral condensate in QED$_3$ are in agreement with
other methods.

\section{QED IN (2+1) DIMENSIONS}

QED$_3$ is a superrenormalizable theory, with a dimensionful coupling:
$e^2$ has dimensions of mass.  This dimensionful parameter plays a role
similar to $\Lambda_{\rm QCD}$ in QCD.  In the chiral limit, it sets
the energy scale.

We use 4-component spinors, such that the fermion mass term is even
under parity.  With one massless fermion, the Hamiltonian exhibits a
global $U(2)$ ``chiral'' symmetry.  A fermion mass term breaks this
symmetry to a $U(1) \times U(1)$ symmetry.  The question is: is this
chiral symmetry broken dynamically?  The order parameter for this
symmetry breaking is the chiral condensate.

There are extensive studies using the Dyson--Schwinger
equations~\cite{dcsb} suggesting that there is dynamical chiral
symmetry breaking in QED$_3$ if the number of fermion flavors is
smaller than some critical number $N_c \sim 3.3$.  However, the scale
of this symmetry breaking (i.e. the magnitude of the condensate) is
rather small.

\subsection{Hamiltonian Lattice Simulations}

Following the notation of Ref.~\cite{Hamer:1997bf}, we consider the
lattice Hamiltonian in (2+1) dimensions
\begin{eqnarray}
  H &=& \frac{g^2}{2a} (W_0 + y W_1 + y^2 W_2)  \; ,
\end{eqnarray}
with
\begin{eqnarray}
  W_0 &=& \sum_l E_l^2 - \mu \sum_{\vec{r}} 
	(-1)^{r_1+r_2} \chi^\dagger(\vec{r}) \chi(\vec{r}) \; ,
\\
  W_1 &=& \sum_{\vec{r},j} \eta_j(\vec{r})
 	\chi^\dagger(\vec{r}) U_j(\vec{r}) \chi(\vec{r}+\hat{j}) 
	+ {\rm h.c.} \; ,
\\
  W_2 &=& -\sum_p\left(U_p + U^\dagger_p\right) \; .
\end{eqnarray}
where $\eta_1(\vec{r}) = (-1)^{r_2+1}$, $\eta_2(\vec{r}) = 1$, 
$y = 1/g^2$, and $U_p$ is the usual plaquette operator.  We use 
a dimensionless mass parameter $\mu = 2 m/e^2$ and a dimensionless
coupling constant $g^2 = e^2 a$ where $a$ is the lattice spacing.  
The continuum limit corresponds to $y \to \infty$ or $g^2 \to 0$.

We use staggered fermions, corresponding to one 4-component fermion
flavor in the continuum limit~\cite{Hamer:1997bf,Burden:qb}.  For
$\mu=0$, the lattice breaks the ``chiral'' $U(2)$ symmetry to a
discrete symmetry generated by a shift of one lattice spacing.  A
nonzero mass term breaks this discrete symmetry.  To study chiral
symmetry breaking on the lattice, we calculate the lattice condensate
$\langle\bar{\psi}\psi\rangle$ in the chiral limit as function of the
lattice coupling $y$.  It is related to the continuum condensate
\begin{eqnarray}
 y^2 \langle\bar{\psi}\psi\rangle^{\hbox{\scriptsize lattice}} 
 	&=& \frac{1}{e^4}\langle\bar{\psi}
		\psi\rangle^{\hbox{\scriptsize continuum}} \; ,
\end{eqnarray}
in the limit $y\rightarrow \infty$.  

We use a finite spatial grid, $N_1 \times N_2$, and calculate
observables 
\begin{eqnarray}
  \langle{\cal O}\rangle &=&
	\frac{\langle\Psi_{\hbox{\scriptsize final}}|
	{\rm e}^{\frac{-H T}{2}} {\cal O}{\rm e}^{\frac{-H t}{2}}
	|\Psi_{\hbox{\scriptsize initial}}\rangle }
	{\langle\Psi_{\hbox{\scriptsize final}}| {\rm e}^{-H t}
	|\Psi_{\hbox{\scriptsize initial}}\rangle }  \; ,
\end{eqnarray}
by dividing the Euclidean time variable $t$ in $N_t$ slices, and
inserting a complete set of states between the time slices.  We choose
a tensor product basis of gauge and fermion states.  We use a standard
Metropolis algorithm to update the gauge field configurations.

For the fermions we use the worldline formalism~\cite{worldline}, in
combination with the loop algorithm~\cite{loop} to update the fermion
configurations.  As initial/final states we choose the strong coupling
($y \to 0$) ground state.  The initial and final fermion states are
identical, up to permutations of fermions.  Numerically, these
permutations of course give rise to the fermion sign problem.  With
the worldline formalism we can keep track of the permutations, and we
use the newly developed {\em zone method}~\cite{dean02} to reduce the
sign problem.

\subsection{Zone method}

\begin{figure}[tbh]
\centerline{\includegraphics[width=5.5cm]{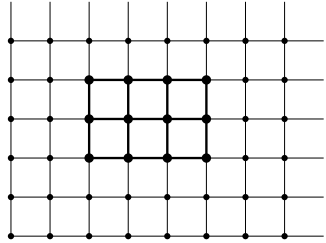}}
\caption{\label{fig:zone}
Illustration of the zone method: an $6\times 8$ spatial lattice
with a $3\times 4$ zone.  The size of this zone is 17, 
as characterized by the number of links inside the sub-lattice.}
\end{figure}

The basic idea~\cite{dean02} is to introduce an $n_1 \times n_2$
spatial sub-lattice (with $n_i \leq N_i$), which we call a {\em zone},
for the fermion simulations, see Fig.~\ref{fig:zone}.  Next, we allow
for permutations of initial and final state fermions {\em inside this
zone} only.
We do not allow for any configuration with fermion permutations
outside the $n_1 \times n_2$ spatial sub-lattice.  Note that at
intermediate time slices the fermions are still allowed to wander
through the entire $N_1 \times N_2$ lattice.  In order to further
reduce the fermion sign problem, we also use a larger value of the
coupling $y'$ and a larger value of the fermion mass $m'$ when the
fermions are outside this zone.

It turns out that observables scale linearly in the zone size,
provided that this zone size is larger than the characteristic
``fermion wandering length'' \cite{dean02}.  For any finite values of
the coupling $y$ and of the time variable $t$, this wandering length
is finite, even for massless fermions.  Thus one can extrapolate the
results from relatively small zones to the entire lattice.

\section{NUMERICAL RESULTS}

There are different ways to define the size of a zone: by number of
lattice points or by the number of links inside the zone.  For large
lattices it does not matter which one one chooses.  However, for the
relatively small lattices we have used sofar, it turns out that the
best way to characterize the zone size is the {\em number of links}
inside the zone.

\begin{figure}[bht]
\includegraphics[width=7.5cm]{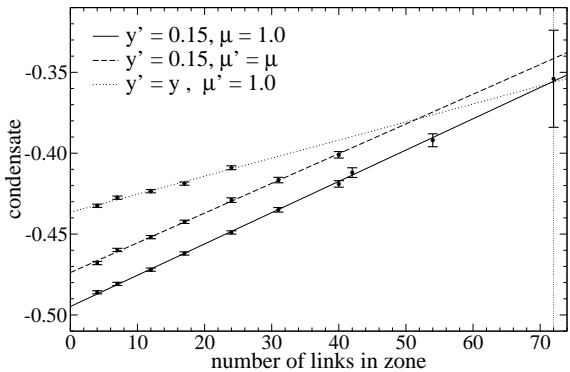}
\caption{\label{fig:results} Numerical results for different zone
sizes within a $6\times 6$ lattice, for fixed values of $y=0.5$, $\mu=0$, 
$t=1.5$, and $N_t=10$.  The total number of links in the lattice 
is 72, where the three linear fits meet (within error bars).}
\end{figure}
In Fig.~\ref{fig:results} we show that, within error bars, the chiral
condensate indeed behaves linearly in the number of links inside the
fermion zone, even for the smallest zones (at least for these
parameter values).  Thus we can use a linear extrapolation to obtain
the lattice condensate for the entire lattice, using three or four
different zone sizes.

\begin{figure}[bht]
\includegraphics[width=7.5cm]{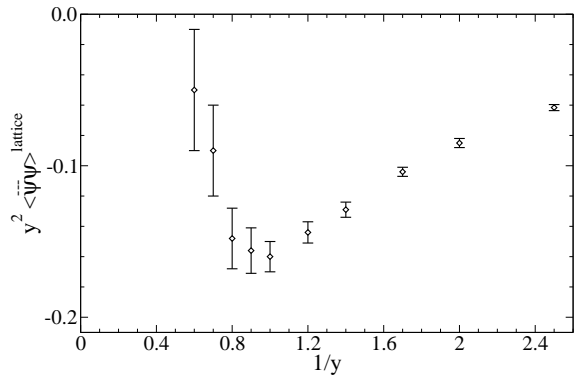}
\caption{\label{fig:final} Our results for the lattice condensate
as function of $y$, obtained by extrapolating several different 
zone sizes for a finite value of $t$.}
\end{figure}
In Fig.~\ref{fig:final} we show our results for a range of values of
the coupling $y$.  These results are obtained on lattices up to
$8\times 8$ using several different zone sizes and for finite values
of $t$.  The error bars are statistical error bars only.  We expect
the finite size effects due to the spatial lattice to be smaller than
these error bars; however, preliminary results indicate the $t =
\infty$ extrapolation could lead to a reduction by about 10\%.  We are
currently working on estimating the finite size effects in more
detail~\cite{LeeMa}.

For $y < 1$ these results agree quite well with the strong coupling
expansion~\cite{Hamer:1997bf}, whereas the strong decrease with
increasing $y$ for $ 1<y<2$ is in good agreement with Euclidean Monte
Carlo simulations~\cite{Burkitt:1987nx}.  It is unclear yet whether or
not the condensate obtained this way is small or exactly zero in the
continuum limit, $y \to \infty$.


\end{document}